\documentclass[12pt,a4paper]{article}
\usepackage{graphicx}
\begin{document}
\textwidth=135mm
 \textheight=200mm
\begin{center}
{\bfseries Double beta decay experiments
\footnote{{\small Talk at the IV International Pontecorvo Neutrino Physics School, 
Alushta, Crimea, Ukraine, 26 September - 6 October,
2010.}}}
\vskip 5mm
A.S. Barabash$^{\dag}$
\vskip 5mm
{\small {\it $^\dag$ Institute of Theoretical and Experimental Physics, 
117218 Moscow, Russia}} \\

\end{center}
\vskip 5mm
\centerline{\bf Abstract}
The present status of double beta decay experiments 
is reviewed. The results of the most sensitive experiments
are discussed. 
Proposals for future double beta 
decay experiments with a sensitivity to the  $\langle m_{\nu} \rangle$ 
at the level of (0.01--0.1) eV are considered.

\vskip 10mm
\section{\label{sec:intro}Introduction}

Interest in neutrinoless double-beta decay has seen a significant renewal in 
recent years after evidence for neutrino oscillations was obtained from the 
results of atmospheric, solar, reactor and accelerator  neutrino 
experiments (see, for example, the discussions in \cite{VAL06,BIL06,MOH06}). 
These results are impressive proof that neutrinos have a non-zero mass. However,
the experiments studying neutrino oscillations are not sensitive to the nature
of the neutrino mass (Dirac or Majorana) and provide no information on the 
absolute scale of the neutrino masses, since such experiments are sensitive 
only to the difference of the masses, $\Delta m^2$. The detection and study 
of $0\nu\beta\beta$ decay may clarify the following problems of neutrino 
physics (see discussions in \cite{PAS03,MOH05,PAS06}):
 (i) lepton number non-conservation, (ii) neutrino nature: 
whether the neutrino is a Dirac or a Majorana particle, (iii) absolute neutrino
 mass scale (a measurement or a limit on $m_1$), (iv) the type of neutrino 
mass hierarchy (normal, inverted, or quasidegenerate), (v) CP violation in 
the lepton sector (measurement of the Majorana CP-violating phases).

Let us consider three main modes of $2\beta$ decay:

\begin{equation}
(A,Z) \rightarrow (A,Z+2) + 2e^{-} + 2\bar { \nu},
\end{equation}

\begin{equation}
(A,Z) \rightarrow (A,Z+2) + 2e^{-},
\end{equation}

\begin{equation}
(A,Z) \rightarrow (A,Z+2) + 2e^{-} + \chi^{0}(+ \chi^{0}).
\end{equation}

The $2\nu\beta\beta$ decay (process (1)) is a second-order
process, which is not forbidden
by any conservation law. The detection of this process provides the
experimental determination  of the 
nuclear matrix  elements (NME) involved
in the double beta decay  processes.  This leads to the development of
theoretical schemes for NME calculations  both in
connection with the $2\nu\beta\beta$ decays as well as the
$0\nu\beta\beta$ decays \cite{ROD06,KOR07,KOR07a,SIM08}.  
Moreover, the study can yield a careful investigation of the
time dependence of the coupling constant for weak interactions
\cite{BAR98,BAR00,BAR03}. 

Recently, it has been pointed out that the $2\nu\beta\beta$ decay allows 
one to investigate particle properties, in particular whether the Pauli 
exclusion principle is violated for neutrinos and thus neutrinos partially obey
Bose--Einstein statistics \cite{DOL05,BAR07b}.

The $0\nu\beta\beta$ decay (process (2)) violates the law
of lepton-number conservation
($\Delta L =2$)
and requires that the Majorana neutrino has a nonzero rest mass. Also, this
process is possible in some
supersymmetric models, where $0\nu\beta\beta$ decay is initiated by
the exchange of supersymmetric
particles. This decay also arises in models featuring an extended
Higgs sector within
electroweak-interaction theory and in some other cases
\cite{KLA98}.

The $0\nu\chi^{0}\beta\beta$ decay (process (3)) requires
the existence of a Majoron. It is a massless
Goldstone boson that arises due to a global breakdown of ($\it{B}$-$\it{L}$)
symmetry, where {\it B} and {\it L} are,
respectively, the baryon and the lepton number. The Majoron, if it
exists, could play a significant role
in the history of the early Universe and in the evolution of
stars. The model of a triplet
Majoron \cite{GEL81} was disproved in 1989
by the data on the decay width of the $Z^{0}$ boson that were
obtained at the LEP accelerator \cite{CAS98}. Despite this, some new models were proposed
\cite{MOH91,BER92}, where $0\nu\chi^{0}\beta\beta$
decay is possible
and where there are no contradictions with the LEP data. A
$2\beta$-decay model that involves the
emission of two Majorons was proposed within supersymmetric
theories \cite{MOH88} and several other models of the
Majoron were proposed in the 1990s. By the term ``Majoron``, one
means massless or light bosons
that are associated with neutrinos. In these models, the Majoron
can carry a lepton charge and is
not required to be a Goldstone boson \cite{BUR93}. A decay process
that involves the emission of two Majoron
is also possible \cite{BAM95}. In models featuring a vector
Majoron, the Majoron is the longitudinal
component of a massive gauge boson emitted in 2$\beta$ decay
\cite{CAR93}. For the sake of simplicity, each such
object is referred to here as a Majoron.
In the Ref. \cite{MOH00}, a ``bulk`` Majoron model was proposed
in the context of the ``brane-bulk`` scenario for particle physics.

The possible two electrons energy spectra for different 2$\beta$
decay modes of $^{100}$Mo are shown
in Fig. 1. Here {\it n} is the spectral 
index, which defines
the shape of the spectrum. For example, for an ordinary Majoron {\it n} = 1, 
for 2$\nu$ decay {\it n} = 5, in the case of a bulk Majoron {\it n} = 2 
and for the process with two Majoron emission {\it n} = 3 or 7.

\begin{figure}
  \begin{center} 
    \includegraphics[height=10cm]{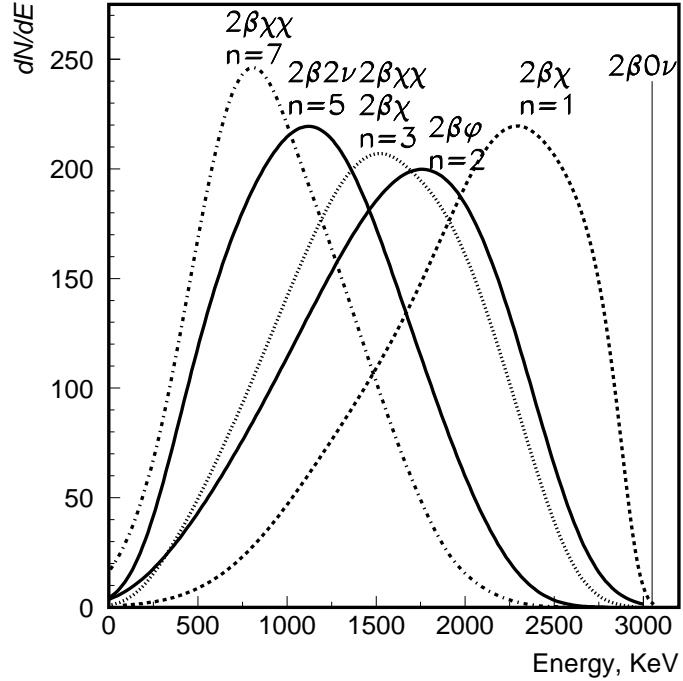} 
  \end{center}
  \caption{Energy spectra of different modes of $2\nu\beta\beta$
$(n=5)$,
$0\nu\chi^{0}\beta\beta$ $(n=1~$,$~2$ and$~3)$ and
$0\nu\chi^{0}\chi^{0}\beta\beta (n=3~$and$~7)$ decays of $~^{100}$Mo.
} 
  \label{figure1}  
\end{figure}

\section{Results of experimental investigations}
The number of possible candidates for double beta decay is quite
large, there are 35 nuclei\footnote{In addition 34
nuclei can undergo double electron
capture, while twenty two nuclei and six nuclei can undergo,
respectively, EC$\beta^{+}$ and 2$\beta^{+}$ decay (see the tables
in \cite{TRE02}).}. However, nuclei for which the double beta transition energy
($E_{2\beta}$) is in excess
of 2 MeV are of greatest interest, since the double beta decay
probability strongly depends on the transition
energy ($\sim E^{11}_{2\beta}$ for $2\nu\beta\beta$ decay, 
$\sim E^{7}_{2\beta}$ for $0\nu\chi^{0}\beta\beta$ decay and 
$\sim E^{5}_{2\beta}$ for $0\nu\beta\beta$ decay). 
In transitions to excited states of the daughter nucleus,
the excitation energy is removed via the
emission of one or more photons, which can be detected, and this
can serve as an additional source
of information about double beta decay. As an example Fig. 2
shows the diagram of energy levels in the
$^{100}$Mo-$^{100}$Tc-$^{100}$Ru nuclear triplet.

\begin{figure}
  \begin{center} 
    \includegraphics[height=10cm]{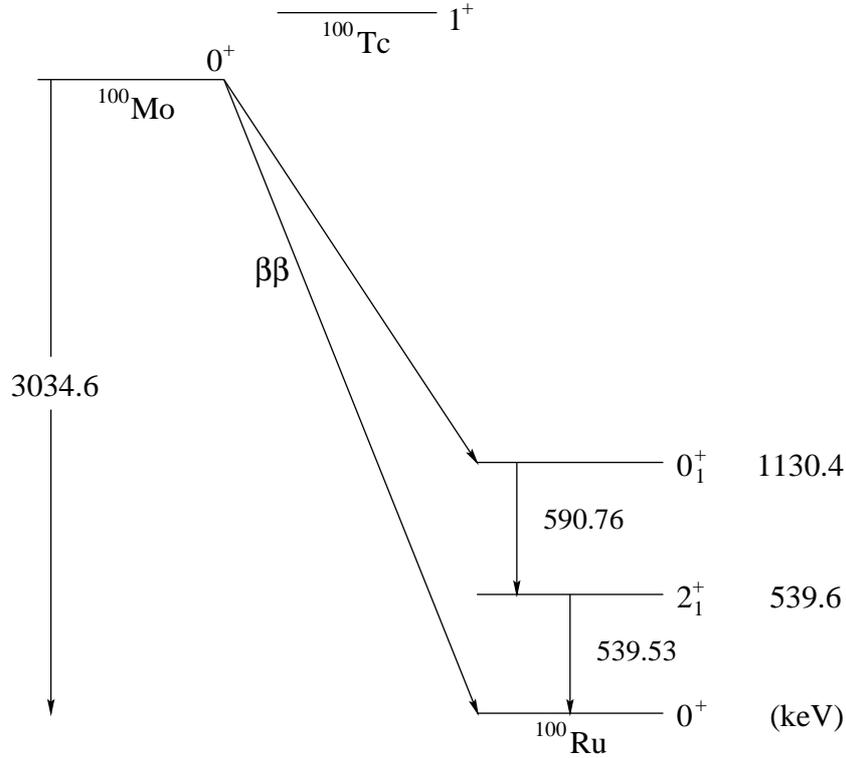} 
  \end{center}
  \caption{Levels scheme for $^{100}$Mo-$^{100}$Tc-$^{100}$Ru. } 
  \label{figure2}  
\end{figure} 

\subsection{Two neutrino double beta decay}

This decay was first recorded in 1950 in a geochemical experiment
with $^{130}$Te \cite{ING49}; in 1967,
$2\nu\beta\beta$ decay was found for $^{82}$Se also in a geochemical
experiment \cite{KIR67}.
Attempts to observe this decay in a direct experiment employing
counters had been futile for
a long time. Only in 1987 could M. Moe, who used a time-projection
chamber (TPC), observe $2\nu\beta\beta$ decay
in $^{82}$Se
for the first time \cite{ELL87}. In the next few years,
experiments were able to detect
$2\nu\beta\beta$
decay in many nuclei. In $^{100}$Mo and $^{150}$Nd 
$2\beta(2\nu)$ decay to the $0^{+}$ excited state of the
daughter nucleus was measured too (see \cite{BAR10}). Also, the
$2\nu\beta\beta$ decay of $^{238}$U was detected in a radiochemical
experiment \cite{TUR91}, and in a geochemical experiment
the ECEC process was detected in $^{130}$Ba (see section 2.4). 
Table 1 displays the present-day averaged and
recommended values of
$T_{1/2}$(2$\nu$) from \cite{BAR10}.
At present, experiments devoted to detecting
$2\nu\beta\beta$ decay are approaching a level
where it is insufficient to just record the
decay. It is necessary to
measure numerous parameters of this process to a high precision 
(energy sum spectrum, single electron energy spectrum and angular 
distribution).
Tracking detectors that are able to record both
the energy of each electron and the angle at which they diverge
are the most appropriate instruments for
solving this problem.

\begin{table}
\caption{Average and recommended $T_{1/2}(2\nu)$ values (from
\cite{BAR10}). }
\bigskip
\begin{tabular}{c|c}
\hline
Isotope & $T_{1/2}(2\nu)$, y \\
\hline
$^{48}$Ca & $4.4^{+0.6}_{-0.5}\times10^{19}$ \\
$^{76}$Ge & $(1.5 \pm 0.1)\times10^{21}$ \\
$^{82}$Se & $(0.92 \pm 0.07)\times10^{20}$ \\
$^{96}$Zr & $(2.3 \pm 0.2)\times10^{19}$ \\
$^{100}$Mo & $(7.1 \pm 0.4)\times10^{18}$ \\
$^{100}$Mo-$^{100}$Ru$(0^{+}_{1})$ & $(5.9^{+0.8}_{-0.6})\times10^{20}$ \\
$^{116}$Cd & $(2.8 \pm 0.2)\times10^{19}$\\
$^{128}$Te & $(1.9 \pm 0.4)\times10^{24}$ \\
$^{130}$Te & $(6.8^{+1.2}_{-1.1}\times10^{20}$ \\
$^{150}$Nd & $(8.2 \pm 0.9)\times10^{18}$ \\
$^{150}$Nd-$^{150}$Sm$(0^{+}_{1})$ & $1.33^{+0.45}_{-0.26}\times10^{20}$
\\
$^{238}$U & $(2.0 \pm 0.6)\times10^{21}$  \\
$^{130}$Ba; ECEC(2$\nu$) & $(2.2 \pm 0.5)\times10^{21}$  \\
\hline
\end{tabular}
\end{table}

\begin{table}
\caption{Best present results on $0\nu\beta\beta$ decay (limits at
90\% C.L.) }
\vspace{0.5cm}
\begin{center}
\begin{tabular*}{\textwidth}{l|@{\extracolsep{\fill}}c|c|c|c}
\hline
Isotope & $T_{1/2}$, y & $\langle m_{\nu} \rangle$, eV & $\langle
m_{\nu} \rangle$, eV & Experiment \\
  ($E_{2\beta}$, keV) & & \cite{ROD06,KOR07,KOR07a,SIM08} & \cite{CAU08}  \\
\hline
$^{76}$Ge (2039) & $>1.9\times10^{25}$ & $<$0.22--0.41 & $<0.69$ & HM
\cite{KLA01} \\
& $\simeq 1.2\times10^{25}$(?) & $\simeq$ 0.28--0.52(?) & $\simeq
0.87(?)$ & Part of HM \cite{KLA04} \\
& $\simeq 2.2\times10^{25}$(?) & $\simeq$ 0.21--0.38(?) & $\simeq
0.64(?)$ & Part of HM \cite{KLA06} \\
& $>1.6\times10^{25}$ & $<$0.24--0.44 & $<0.75$ & IGEX
\cite{AAL02a} \\
\hline
$^{130}$Te (2529) & $>2.8\times10^{24}$ & $<$0.29--0.59 & $<0.77$ &
CUORICINO \cite{AND11} \\
$^{100}$Mo (3034) & $>1.1\times10^{24}$ & $<$0.45--0.93 & - & NEMO-3$^{*)}$ \cite{BAR10a}  \\
$^{136}$Xe (2458) & $>4.5\times10^{23**)}$ & $<$1.14--2.68 & $<2.2$ & DAMA
\cite{BER02} \\
$^{82}$Se (2995) & $>3.6\times10^{23}$ & $<$0.89--1.61 & $<2.3$ & NEMO-3$^{*)}$ \cite{BAR10a}  \\
$^{116}$Cd  (2805) & $>1.7\times10^{23}$ & $<$1.40--2.76 & $<1.8^{***)}$ &
SOLOTVINO \cite{DAN03} \\

\hline
\end{tabular*}
\flushleft{
$^{*)}$ Current experiments; $^{**)}$ conservative limit from \cite{BER02} is presented;
$^{***)}$ NME from \cite{CAU07} is used.}
\end{center}
\end{table}

\subsection{Neutrinoless double beta decay}

In contrast to two-neutrino decay, neutrinoless double beta decay
has not yet been observed\footnote{The possible
exception is the result with $^{76}$Ge, published by a fraction of
the  Heidelberg-Moscow
Collaboration, $T_{1/2} \simeq 1.2\times 10^{25}$ y \cite{KLA04} or 
$T_{1/2} \simeq 2.2\times 10^{25}$ y \cite{KLA06}. First time the 
``positive'' result was mentioned in \cite{KLA01a}. 
The Moscow part of the
Collaboration does not agree with this conclusion \cite{BAK03} and
there are others who are critical
of this result \cite{AAL02,ZDE02,STR05}. Thus, at the present time, this
``positive'' result is not accepted by the
``2$\beta$ decay community'' and it has to be checked by new
experiments.}, although it is easier to detect it. In
this case, one seeks, in the
experimental spectrum, a peak of energy equal to the double beta
transition energy and of width determined
by the detector's resolution.

The constraints on the existence of
$0\nu\beta\beta$ decay are presented in Table 2 for the nuclei
that are the most promising candidates. In calculating constraints
on $\langle m_{\nu} \rangle$, the
nuclear matrix elements from \cite{ROD06,KOR07,KOR07a,SIM08} were used (3-d column). It
is advisable to employ the calculations from
these studies, because the calculations are the most thorough and
take into account the most recent theoretical
achievements. In these papers $g_{pp}$ values ($g_{pp}$ is 
parameter of the QRPA theory) were fixed using experimental 
half-life values for $2\nu$ decay and then
NME(0$\nu$) were calculated. 
In column four, limits on $\langle m_{\nu} \rangle$,
which were obtained using the NMEs
from a recent Shell Model (SM) calculations \cite{CAU08}, are presented (for $^{116}$Cd NME 
from \cite{CAU07} is used).

From Table 2 using NME values 
from \cite{ROD06,KOR07,KOR07a,SIM08}, the limits on 
$\langle m_{\nu} \rangle$ for $^{130}$Te are comparable 
with the $^{76}$Ge results. Now one cannot 
select any experiment as the best one. 
The assemblage of sensitive experiments
for different nuclei permits one to increase the reliability of the limit 
on $\langle m_{\nu} \rangle$. Present conservative limit can be set as 0.75 eV.

\subsection{Double beta decay with Majoron emission}

Table 3 displays the best present-day constraints for an "ordinary"
 Majoron ({\it n} = 1). The NME from the 
following works were used, 3-d column: \cite{ROD06,KOR07,KOR07a,SIM08}, 4-th 
column: \cite{CAU08}.
The "nonstandard" models of the Majoron were experimentally tested
in \cite{GUN97} for $^{76}$Ge
and in \cite{ARN00} for $^{100}$Mo,
$^{116}$Cd, $^{82}$Se, and $^{96}$Zr. Constraints on the decay
modes involving the emission of two Majorons were also
obtained for $^{100}$Mo \cite{TAN93}, $^{116}$Cd \cite{DAN03}, and
$^{130}$Te \cite{ARN03}. In a recent NEMO Collaboration papers, 
new results for these processes in $^{100}$Mo \cite{ARN06}, $^{82}$Se \cite{ARN06},
$^{150}$Nd \cite{ARG09} and $^{96}$Zr \cite{ARN10}
were obtained with the NEMO-3 detector. 
Table 4 gives the best experimental
constraints on decays accompanied by the emission of one or two
Majorons (for {\it n} = 2, 3, and 7).
Hence at the present time only limits on double beta decay with
Majoron emission have been obtained (see Tables 3 and 4).
A conservative present limit on the coupling constant of ordinary 
Majoron to the
neutrino is $\langle g_{ee} \rangle < 1.9 \times 10^{-4}$.

\begin{table}
\caption{Best present limits on $0\nu\chi^{0}\beta\beta$ decay
(ordinary Majoron) at 90\% C.L.}
\vspace{0.5cm}
\begin{center}
\begin{tabular*}{\textwidth}{l|@{\extracolsep{\fill}}c|c|c}
\hline
Isotope  & $T_{1/2}$, y & $\langle g_{ee} \rangle$, \cite{ROD06,KOR07,KOR07a,SIM08}
 & $\langle
g_{ee} \rangle$, \cite{CAU08} \\  
 ($E_{2\beta}$, keV)&  &  & \\ 
\hline
$^{76}$Ge (2039) & $>6.4\times10^{22}$ \cite{KLA01} & $<$ (0.54--1.44)$\times10^{-
4}$ & $<2.4\times10^{-4}$ \\
$^{82}$Se (2995) & $>1.5\times10^{22}$ \cite{ARN06} & $<$ (0.58--1.19)$\times10^{-4}$ & 
$<1.9\times10^{-4}$ \\
$^{100}$Mo (3034) & $>2.7\times10^{22}$ \cite{ARN06} & $<$ (0.35--0.85)$\times10^{-4}$ & - \\
$^{116}$Cd (2805) & $>8\times10^{21}$ \cite{DAN03} & $<$ (0.79--2.56)$\times10^{-
4}$ & $<1.7\times10^{-4}$$^{**)}$ \\
$^{128}$Te (867) & $>1.5\times10^{24}$(geochem)& $<$ (0.63--1)$\times10^{-4}$ & 
$<1.4\times10^{-4}$ \\
 & \cite{MAN91,BAR10}  &   & \\
$^{136}$Xe (2458) & $>1.6\times10^{22*)}$ \cite{BER02} & $<$ (1.51--3.54)$\times10^{-
4}$ & $<2.9\times10^{-4}$ \\
\hline
\end{tabular*}
\end{center}
\flushleft {$^{*)}$ Conservative limit from \cite{BER02} is presented; 
$^{**)}$ NME from \cite{CAU07} is used.}
\end{table}

\begin{table}
\caption{Best present limits on $T_{1/2}$ for decay with one and
two Majorons at 90\% C.L. for modes
with spectral index {\it n} = 2, {\it n} = 3 and {\it n} = 7}
\vspace{0.5cm}
\begin{center}
\begin{tabular*}{\textwidth}{l|@{\extracolsep{\fill}}c|c|c}
\hline
Isotope ($E_{2\beta}$, keV) & {\it n} = 2 & {\it n} = 3 &  {\it n} = 7  \\
\hline
$^{76}$Ge (2039) & - & $>5.8\times10^{21}$ \cite{GUN97} &
$>6.6\times10^{21}$ \cite{GUN97} \\
$^{82}$Se (2995) & $>6\times10^{21}$ \cite{ARN06} & $>3.1\times10^{21}$
\cite{ARN06} & $>5\times10^{20}$ \cite{ARN06} \\
$^{96}$Zr (3350) & $>9.9\times10^{20}$ \cite{ARN10} & $>5.8\times10^{20}$ \cite{ARN10} &
$>1.1\times10^{20}$ \cite{ARN10} \\
$^{100}$Mo (3034) & $>1.7\times10^{22}$ \cite{ARN06} & $>1\times10^{22}$
\cite{ARN06} & $>7\times10^{19}$ \cite{ARN06} \\
$^{116}$Cd (2805) & $>1.7\times10^{21}$ \cite{DAN03} & $>8\times10^{20}$
\cite{DAN03} & $>3.1\times10^{19}$ \cite{DAN03} \\
$^{130}$Te (2529) & - & $>9\times10^{20}$ \cite{ARN03} & - \\
$^{128}$Te (867) & $>1.5\times10^{24}$  & $>1.5\times10^{24}$  
& $>1.5\times10^{24}$ \\
(geochem)  & \cite{MAN91,BAR10} & \cite{MAN91,BAR10} & \cite{MAN91,BAR10}\\
$^{150}$Nd (3371) & $>5.4\times10^{20}$ \cite{ARG09} & $>2.2\times10^{20}$ \cite{ARG09} &
$>4.7\times10^{19}$ \cite{ARG09} \\
\hline
\end{tabular*}
\end{center}
\end{table}

\subsection{2$\beta^+$, EC$\beta^+$, and ECEC processes}

Much less attention has been given to the investigation of $2\beta^+$, $\beta^+$EC and ECEC 
processes although such attempts were done from time to time in the past (see review 
\cite{BAR10b}). Again, the main interest here is connected with neutrinoless decay:

\begin{equation}
(A,Z) \rightarrow (A,Z-2) + 2e^{+}, 
\end{equation}

\begin{equation}
e^- + (A,Z) \rightarrow (A,Z-2) + e^{+} + X,
\end{equation}

\begin{equation}
e^- + e^- + (A,Z) \rightarrow (A,Z-2)^* \rightarrow (A,Z-2)+\gamma + 2X.
\end{equation}

There are 34 candidates for these processes. 
Only 6 nuclei can undergo all the above mentioned processes and 16 nuclei can undergo $\beta^+$EC 
and ECEC while 12 can undergo only ECEC. Detection of the neutrinoless mode in the 
above processes enable one to determine the effective Majorana neutrino mass 
$\left<m_\nu\right>$, parameters of right-handed current admixture in electroweak 
interaction ($\left<\lambda\right>$ and $\left<\eta\right>$), etc. 

Process (4) has a very nice signature because, in addition to two positrons, four annihilation 511 
keV gamma quanta will be detected. On the other hand, the rate for this process should be much 
lower in comparison with $0\nu\beta\beta$ decay because of substantially lower kinetic energy 
available in such a transition (2.044 MeV is spent for creation of two positrons) and of the 
Coulomb barrier for positrons.  There are only six candidates for this type of decay: $^{78}$Kr, 
$^{96}$Ru, $^{106}$Cd, $^{124}$Xe, $^{130}$Ba and $^{136}$Ce. The half-lives of most prospective 
isotopes are estimated to be $\sim 10^{27}$--$10^{28}$ y (for $\left<m_\nu\right> = 1$ eV) 
\cite{SUH03,HIR94}; this is approximately $10^3$--$10^4$ times higher than for $0\nu\beta\beta$ decay for 
such nuclei as $^{76}$Ge, $^{100}$Mo, $^{82}$Se and $^{130}$Te.

Process (5) has a nice signature (positron and two annihilation 511 keV gammas) and is not as 
strongly suppressed as $2\beta^+$ decay. In this case, half-life estimates for the best nuclei 
give $\sim 10^{26}$--$10^{27}$ y (again for $\left<m_\nu\right> = 1$ eV) \cite{SUH03,HIR94}.

In the last case (process (6)), the atom de-excites emitting two X-rays and the nucleus de-excites 
emitting one $\gamma$-ray (bremsstrahlung photon)\footnote{In fact the processes with irradiation 
of inner conversion electron, $e^+e^-$ pair or two gammas are also possible \cite{DOI93} (in 
addition, see discussion in \cite{SUJ04}). These possibilities are especially important in 
the case of the ECEC$(0\nu)$ transition with the capture of two electrons from the {\it K} shell. In this case 
the transition with irradiation of one $\gamma$ is strongly suppressed \cite{DOI93}.}.
 For a transition to an excited state of the daughter nucleus, besides a bremsstrahlung photon, 
$\gamma$-rays are emitted from the decay of the excited state.  Thus, there is a clear 
signature for this process. The rate is practically independent of decay 
energy and increases with both decreasing bremsstrahlung photon energy and increasing {\it Z} 
\cite{SUJ04,VER83}.  The rate is quite low even for heavy nuclei, with $T_{1/2} \sim 10^{28} 
-10^{31}$  y ($\left<m_\nu\right> = 1$ eV) \cite{SUJ04}. The rate can be increased 
in $\sim$ 10$^6$ times if resonance conditions 
exist (see Section 2.4.1). 

For completeness, let us present the two-neutrino modes of $2\beta^+$, $\beta^+$EC and ECEC 
processes:

\begin{equation}
(A,Z) \rightarrow (A,Z-2) + 2e^{+} +2\nu,
\end{equation}

\begin{equation}
e^- + (A,Z) \rightarrow (A,Z-2) + e^{+} + 2\nu + X,
\end{equation}

\begin{equation}
e^- + e^- + (A,Z)  \rightarrow (A,Z-2)+ 2\nu + 2X.
\end{equation}

These processes are not forbidden by any conservation laws, and their observation is interesting 
from the point of view of investigating nuclear-physics aspects of double-beta decay. Processes 
(7) and (8) are quite strongly suppressed because of low phase-space volume, and investigation of 
process (9) is very difficult because one only has low energy X-rays to detect.  In the case of 
double-electron capture, it is again interesting to search for transitions to the excited states 
of daughter nuclei, which are easier to detect experimentally \cite{BAR94}.
For the best candidates half-life is estimated as $\sim 10^{27}$ y for $\beta^+\beta^+$,
$\sim 10^{22}$ y for $\beta^+$EC and $\sim 10^{21}$ y for ECEC process 
\cite{HIR94}. 

During the last few years, interest in the $\beta^+\beta^+$, $\beta^+$EC and ECEC processes 
has greatly increased. For the first time a positive result was obtained in a geochemical 
experiment with $^{130}$Ba, where the ECEC$(2\nu)$ process was detected with a 
half-life of $(2.2 \pm 0.5)\times 10^{21}$ y \cite{MES01}. Recently new  
limits on the ECEC($2\nu$) process in the promising candidate isotopes ($^{78}$Kr and $^{106}$Cd) 
were established 
($2.4\times 10^{21}$ y \cite{GAV10} and 
$4.1\times 10^{20}$ y \cite{RUK10}, respectively).
Very recently  
$\beta^+$EC and ECEC processes in $^{120}$Te \cite{WIL06,BAR07d}, $^{74}$Se \cite{BAR07e},
$^{64}$Zn \cite{KIM07,BEL08} and $^{112}$Sn \cite{KIM07,BAR08,ZUB08} 
were investigated.  Among the recent papers there are a few new theoretical 
papers with half-life estimations \cite{SUH03,DOM05,CHA05,RAI06,SHU07,MIS08,RAT09,SHU09}. Nevertheless the 
$\beta^+\beta^+$, $\beta^+$EC and ECEC processes have  not been investigated very well theoretically 
or experimentally. One can imagine some unexpected results here, which is why any 
improvements in experimental sensitivity for such transitions has merit. 

Table 5 gives a compendium of the best present-day constrains for 2$\beta^+$, EC$\beta^+$, and ECEC processes
and the result of the geochemical experiment that employed $^{130}$Ba and which yields the first 
indication of the observation of ECEC(2$\nu$) capture.

\begin{table}
\caption{Most significant experimental results for 2$\beta^+$, EC$\beta^+$, and ECEC processes 
(all limits are presented at a 90\% C.L.). Here {\it Q} is equal to $\Delta M$ (atomic mass 
difference of parent and daughter nuclei) for ECEC, ($\Delta M$ - 1022 keV) for EC$\beta^+$ 
and ($\Delta M$ - 2044 keV) for 2$\beta^+$.}
\begin{center}
\begin{tabular*}{\textwidth}{l|@{\extracolsep{\fill}}c|c|c|c}
\hline
 Decay type & Nucleus & {\it Q}, keV & $T_{1/2}$, y & References \\
\hline 
ECEC(0$\nu$) & $^{130}$Ba & 2611 &$> 4\times10^{21}$ & \cite{BAR96a} \\
             & $^{78}$Kr & 2866 &$> 2.4\times10^{21}$ & \cite{GAV10} \\  
              & $^{132}$Ba & 839.9 &$> 3\times10^{20}$ & \cite{BAR96a} \\    
              & $^{106}$Cd & 2771 &$> 1.6\times10^{20}$ & \cite{RUK10} \\    
 ECEC(2$\nu$) & $^{130}$Ba & 2611 &$> 4\times10^{21}$ & \cite{BAR96a} \\
              &            & &$ = 2.1^{+3.0}_{-0.8}\times10^{21}$ & \cite{BAR96a} \\
              &            & &$= (2.2 \pm 0.5)\times10^{21}$ & \cite{MES01} \\
              & $^{78}$Kr & 2866 &$> 2.4\times10^{21}$ & \cite{GAV10} \\ 
              & $^{106}$Cd & 2771 &$> 4.1\times10^{20}$ & \cite{RUK10} \\
              & $^{132}$Ba & 839.9 &$> 3\times10^{20}$ & \cite{BAR96a} \\    
 EC$\beta^+(0\nu)$  & $^{130}$Ba & 1589 &$> 4\times10^{21}$ & \cite{BAR96a} \\  
              & $^{78}$Kr & 1844 &$> 2.5\times10^{21}$ & \cite{SAE94} \\  
              & $^{58}$Ni & 903.8 &$> 4.4\times10^{20}$ & \cite{VAS93} \\  
              & $^{106}$Cd & 1749 & $> 3.7\times10^{20}$ & \cite{BEL99} \\  
              & $^{92}$Mo & 627.1 &$> 1.9\times10^{20}$ & \cite{BAR97} \\  
 EC$\beta^+(2\nu)$  & $^{130}$Ba & 1589 &$> 4\times10^{21}$ & \cite{BAR96a} \\               
              & $^{58}$Ni & 903.8 & $> 4.4\times10^{20}$ & \cite{VAS93} \\  
              & $^{106}$Cd & 1749 & $> 4.1\times10^{20}$ & \cite{BEL99} \\  
              & $^{92}$Mo & 627.1 & $> 1.9\times10^{20}$ & \cite{BAR97} \\  
              & $^{78}$Kr & 1844 & $> 7\times10^{19}$ & \cite{SAE94} \\  
 2$\beta^+(0\nu)$ & $^{130}$Ba & 567 & $> 4\times10^{21}$ & \cite{BAR96a} \\    
              & $^{78}$Kr & 822 & $> 1\times10^{21}$ & \cite{SAE94} \\  
              & $^{106}$Cd & 727 & $> 2.4\times10^{20}$ & \cite{BEL99} \\  
 2$\beta^+(2\nu)$ & $^{130}$Ba & 567 & $> 4\times10^{21}$ & \cite{BAR96a} \\    
              & $^{78}$Kr & 822 & $> 1\times10^{21}$ & \cite{SAE94} \\  
              & $^{106}$Cd & 727 & $> 2.4\times10^{20}$ & \cite{BEL99} \\  
\hline
\end{tabular*}
\end{center}
\end{table}

\subsubsection{ECEC(0$\nu$) resonance transition to the excited states}

In \cite{WIN55} it was the first mentioned that in the case of ECEC(0$\nu$)
transition a resonance condition could exist for transitions to a ``right energy'' excited state 
of the daughter nucleus, when the decay energy is closed to zero.
In 1982 the same idea was proposed for transitions to the ground state \cite{VOL82}. 
In 1983 this transition was discussed for $^{112}$Sn-$^{112}$Cd ($0^+$; 1871 keV)  
\cite{BER83}. In 2004 the idea 
was reanalyzed in \cite{SUJ04} and new resonance condition for the decay was formulated. 
The possible enhancement of the transition rate was estimated as $\sim$ 10$^6$ \cite{BER83,SUJ04}, 
which means that the process starts to be
competitive with $0\nu\beta\beta$ decay sensitivity to neutrino mass
and it is possible to check this by experiment.
There are several candidates for such resonance transitions, to the ground ($^{152}$Gd, $^{164}$Eu and 
$^{180}$W) and to the excited states ($^{74}$Se, $^{78}$Kr, $^{96}$Ru, $^{106}$Cd, $^{112}$Sn, 
$^{130}$Ba, $^{136}$Ce and $^{162}$Er) of daughter nuclei.
The precision needed to realize resonance conditions is well below 1 keV. To select the best 
candidate from the above list one will have to know the atomic mass difference with an 
accuracy better than 1 keV.
Unfortunately by the moment for all mentioned above isotopes the accuracy 
is mainly on the level 2-13 keV. To select the best candidate from the above list one will have 
to know the atomic mass difference with an accuracy better than 1 keV.
In fact, it is possible to know these values with much better accuracy 
and recently the atomic-mass difference between $^{112}$Sn and $^{112}$Cd 
was measured with accuracy 
0.16 keV \cite{RAH09} and between $^{74}$Se and $^{74}$Ge with accuracy 
0.007 keV \cite{MOU10} and 0.049 keV \cite{KOL10}\footnote{Unfortunately
these measurements demonstrate that the strong enhancement scenario for the 
$^{112}$Sn and $^{74}$Se decays is disfavored.}. 
The experimental search
for such a resonance transition in $^{74}$Se to $^{74}$Ge ($2^+$; 1206.9 keV) was
performed yielding a limit $T_{1/2} > 5.5\times10^{18}$ y \cite{BAR07e}. 
Recently the limits
on the level of $1.6\times10^{20}$ yr, $(0.5-1.3)\times10^{19}$ yr and 
$\sim (2-4)\times10^{15}$ yr for the resonant neutrinoless transitions in $^{106}$Cd \cite{RUK10}, 
$^{96}$Ru \cite{BEL09a} and $^{136}$Ce \cite{BEL09b} were obtained.
Resonance transition in $^{112}$Sn was investigated by three different experimental groups using 
natural and enriched samples of tin \cite{BAR08,ZUB08,DAW08,KID08,BAR09}. The more strong 
limit of $T_{1/2} > 4.7\times10^{20}$ yr was obtained for the transition
to the $0^+$ state at 1871 keV with the 53 g enriched tin sample\cite{BAR09}.
It has also been demonstrated that using enriched $^{112}$Sn (or $^{74}$Se) at an installation 
such as GERDA or MAJORANA
a sensitivity on the level $\sim 10^{26}$ y can be reached.
The best present limits are presented in Table 6.

\begin{table}
\caption{Best present limits on ECEC(0$\nu$) to the excited state 
at a 90\% C.L. for isotope-candidates with possible resonance enhancement.
Here $\Delta$ M is the atomic mass 
difference of parent and daughter nuclei, $E^*(J^{\pi})$ is the energy of the excited 
state of the daughter nuclide (with its spin and parity in parenthesis).
}
\begin{center}
\begin{tabular}{l|c|c|c|c}
\hline
 Nucleus & Abund., $\%$ & $\Delta$ M, keV & $E^*(J^{\pi})$ & $T_{1/2}$, y \\
\hline 
$^{74}$Se & 0.89 & $1209.240 \pm 0.007$ & 1204.2 ($2^+$) & $> 5.5\times10^{18}$ \cite{BAR07e} \\
$^{78}$Kr & 0.35 & $2846.4 \pm 2.0$ & 2838.9 ($2^+$) & $> 1\times10^{21 *)}$  \\
$^{96}$Ru & 5.52 & $2718.5 \pm 8.2$ & 2700.2 ($2^+$) & $> 4.9\times10^{18}$ \cite{BEL09a} \\            
&  &  & 2712.68.1 (?) & $> 1.3\times10^{19}$ \cite{BEL09a}  \\
$^{106}$Cd & 1.25 & $2770 \pm 7.2$ & 2741.0 ($4^+$) & $> 1.6\times10^{20}$ \cite{RUK10}  \\
&   &   & 2748.2 (2,$3^-$)  & -  \\
$^{112}$Sn & 0.97 & $1919.82 \pm 0.16$ & 1871.0 ($0^+$) & $> 4.7\times10^{20}$ \cite{BAR09} \\
$^{130}$Ba & 0.11 & $2617.1 \pm 2.0$ & 2608.4 (?) & $> 1.5\times10^{21 **)}$  \\
&  &  & 2544.43 (?) & $> 1.5\times10^{21 **)}$ \\
$^{136}$Ce & 0.20 & $2418.9 \pm 13$ & 2399.9 ($1^+$,$2^+$) & $> 4.1\times10^{15}$ \cite{BEL09b} \\
 &  &  & 2392.1 ($1^+$,$2^+$) & $> 2.4\times10^{15}$ \cite{BEL09b}\\
$^{162}$Er & 0.14 & $1843.8 \pm 5.6$ & 1745.7 ($1^+$) & - \\
&  &  & 1782.68 ($2^+$) & - \\
\hline
\end{tabular}
\flushleft{$^{*)}$ Extracted from results for the ECEC(2$\nu;0^+$-$0^+_{g.s.}$) transition obtained 
for $^{78}$Kr \cite{GAV10};
$^{**)}$ extracted from geochemical experiments \cite{BAR96a,MES01}.}
\end{center}
\end{table}

\section{Large-scale current experiment NEMO-3 \cite{ARN05,ARN05a,ARN04}}



This tracking experiment, in contrast to experiments
with $^{76}$Ge, detects not only the total energy deposition,
but other parameters of the process, including the energy of
the individual electrons, angle between them,
and the coordinates of the event in the source plane. The
performance of the detector was studied with the NEMO-2
prototype \cite{ARN95}. Since June of 2002, the NEMO-3 detector
has operated in the Frejus Underground Laboratory (France)
located at a depth of 4800 m w.e. The detector has a cylindrical
structure and consists of 20 identical sectors
 (see Fig.3).
A thin (30--60 mg/cm$^{2}$) source containing double beta-decaying
nuclei and natural material foils have a total area of 20 m$^{2}$ and a weight
of up
to 10 kg was placed in the detector. The basic principles of
detection are identical to those used in the NEMO-2
detector. The energy of the electrons is measured by plastic
scintillators (1940 individual counters), while the tracks
are reconstructed on the basis of information obtained in the
planes of Geiger cells (6180 cells) surrounding the source
on both sides. The tracking volume of the detector is filled with
a mixture consisting of $\sim$ 95\% He, 4\% alcohol, 1\% Ar
and 0.1\% water at slightly above atmospheric
pressure. In addition, a magnetic field with a strength of 25 G
parallel to the detector's axis is created by a solenoid
surrounding the detector. The magnetic field is used to identify
electron-positron pairs so as to suppress this
source of background.

\begin{figure}
  \begin{center} 
    \includegraphics[height=10cm]{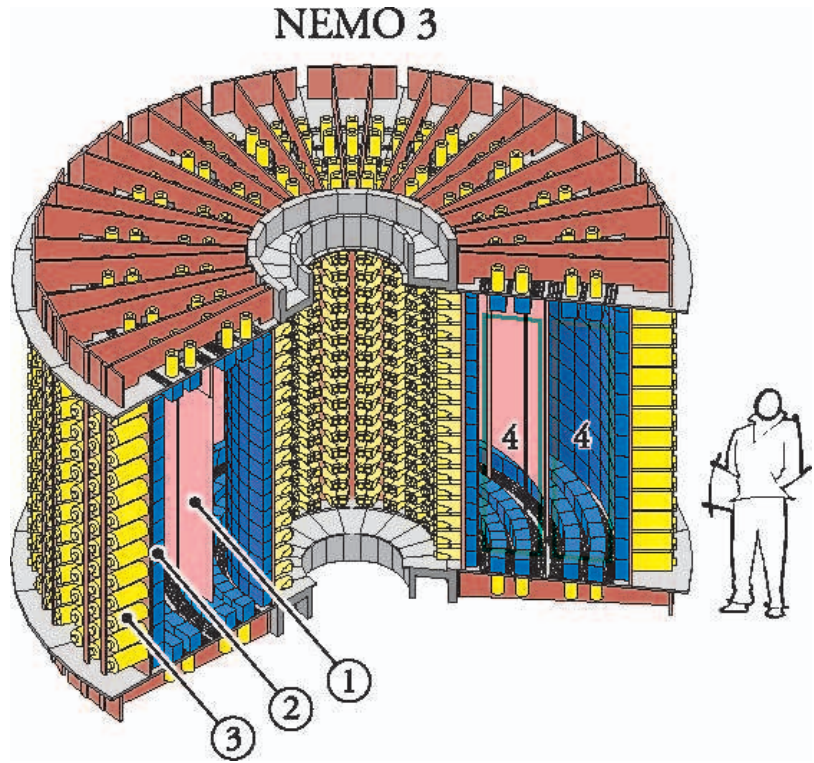} 
  \end{center}
  \caption{The NEMO-3 detector without shielding \cite{ARN05a}: 1 -- source foil;
2 -- plastic scintillator; 3 -- low radioactivity PMT; 4 --
tracking chamber.} 
  \label{figure3}  
\end{figure}

The main characteristics of the detector are the following. The
energy resolution of the scintillation counters lies in
the interval 14--17\% FWHM for electrons of energy 1 MeV. The time
resolution is 250 ps for an electron energy of
1 MeV and the accuracy in reconstructing the vertex of 2{\it e}$^{-
}$ events is 1 cm.
The detector is surrounded by a passive shield consisting of 20 cm
of steel and 30 cm of borated water. The level of
radioactive impurities in structural materials of the detector and
of the passive shield was tested in measurements
with low-background HPGe detectors.

Measurements with the NEMO-3 detector revealed that tracking
information, combined with time and energy measurements,
makes it possible to suppress the background efficiently. That
NEMO-3 can be used to investigate almost all isotopes
of interest is a distinctive feature of this facility. At the
present time, such investigations are being performed
for seven isotopes; these are $^{100}$Mo, $^{82}$Se, $^{116}$Cd,
$^{150}$Nd, $^{96}$Zr, $^{130}$Te, and $^{48}$Ca (see
Table 7). As mentioned above, foils of copper and natural (not
enriched) tellurium were placed in the detector to perform
background measurements.

\begin{figure}
\begin{center}
\includegraphics[scale=0.4]{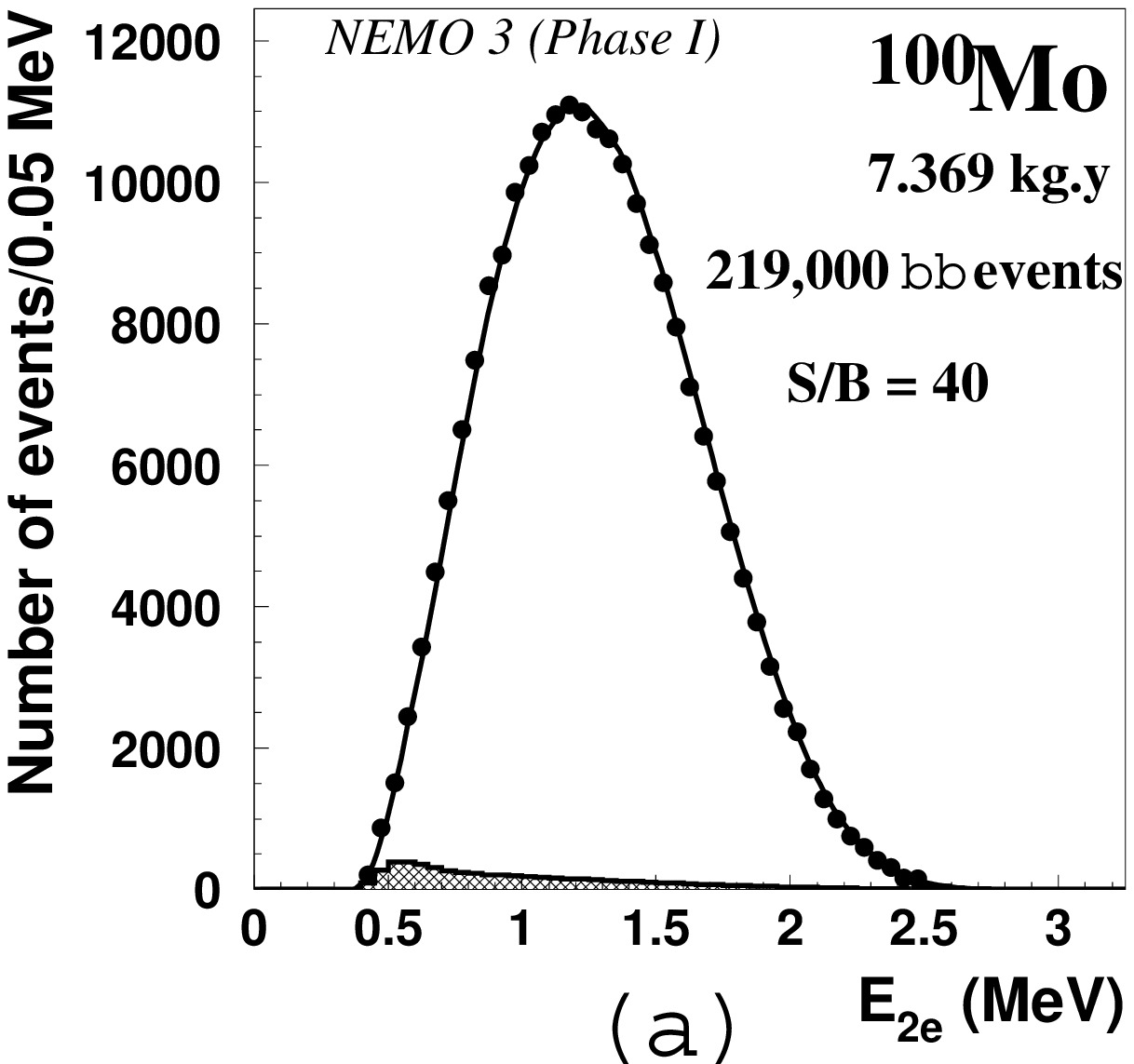}
\includegraphics[scale=0.4]{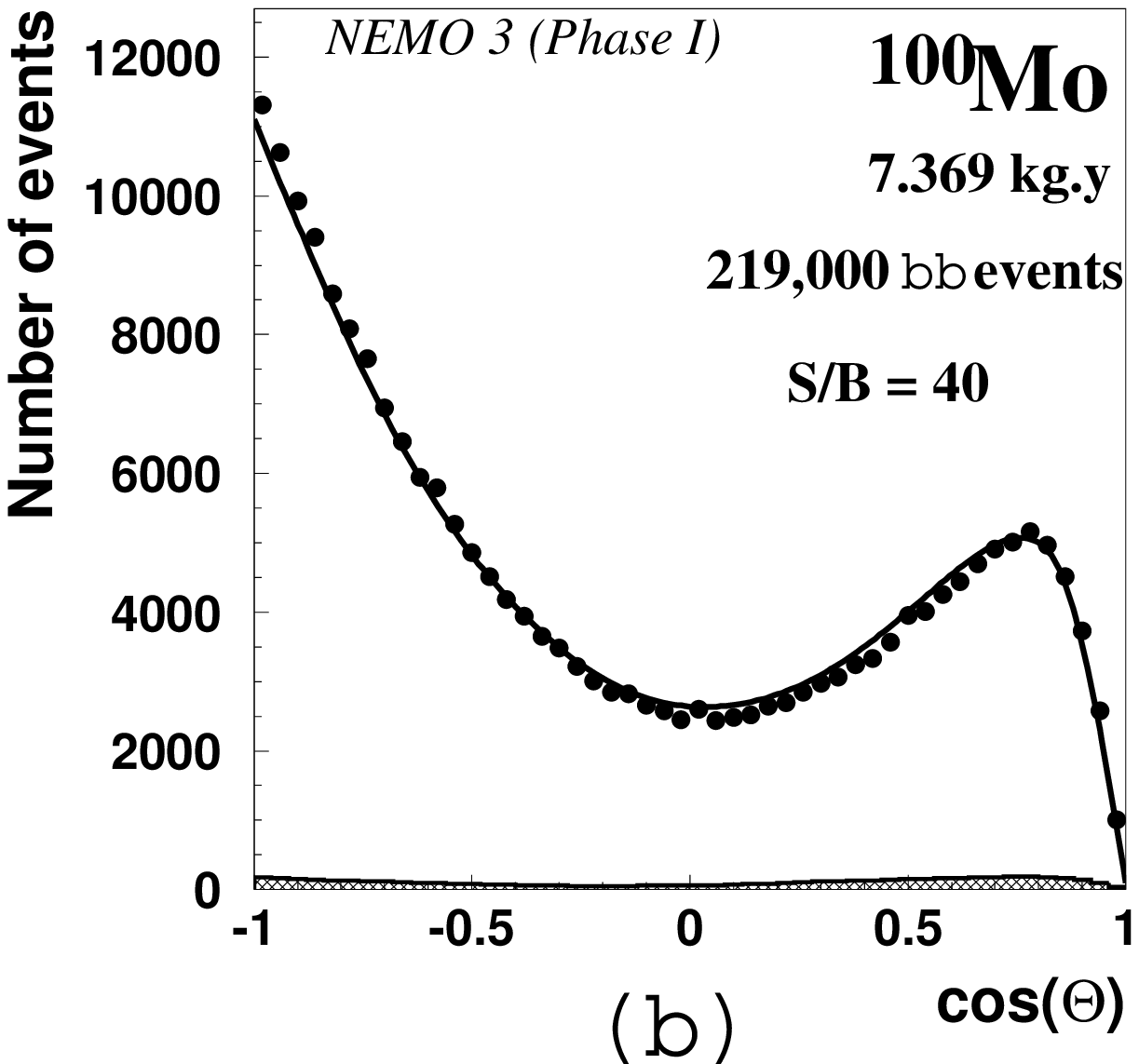}
\includegraphics[scale=0.4]{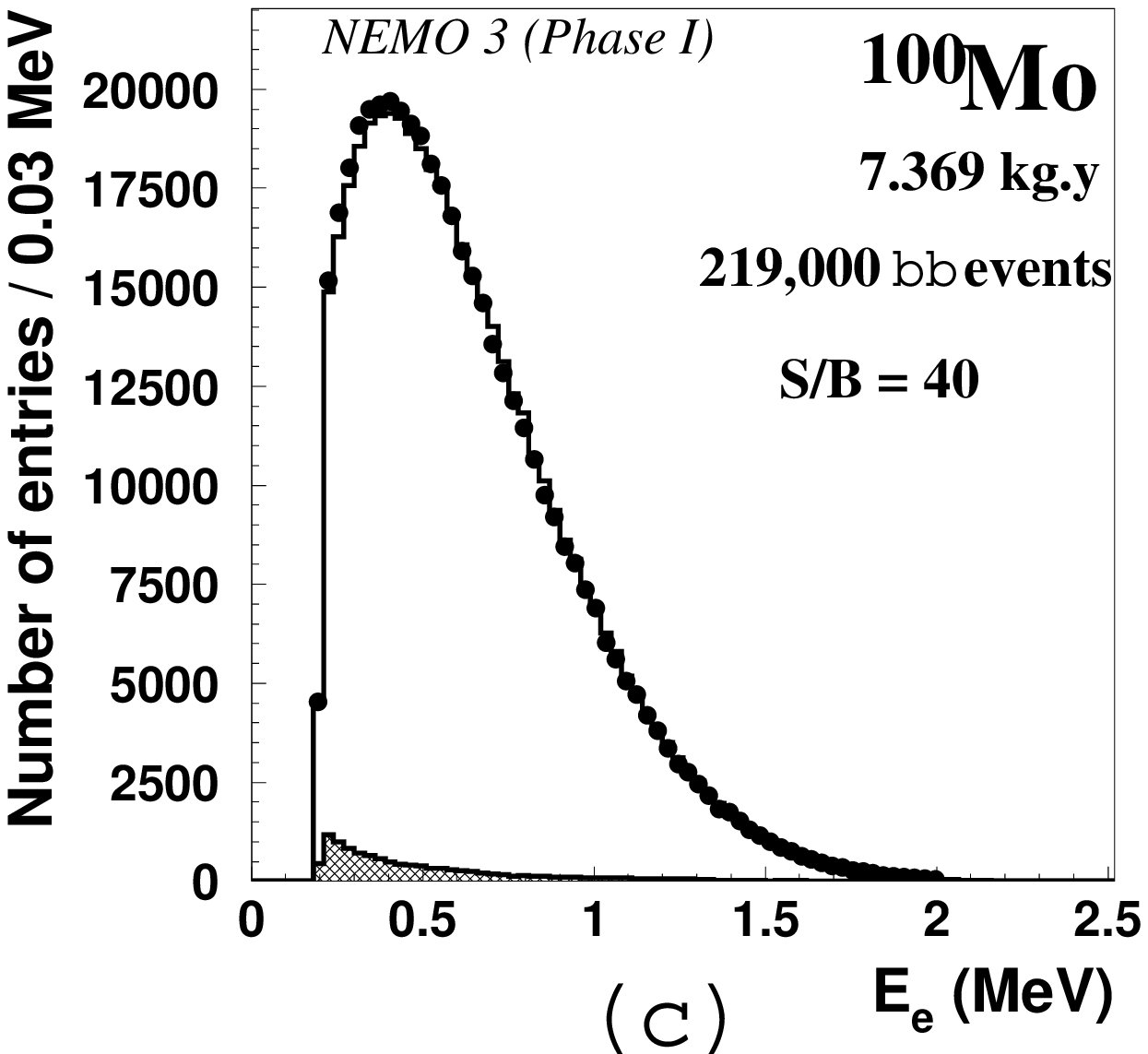}
\caption{\label{fig:figure4} Energy sum spectrum of the two
electrons ({\it a}), angular
distribution of the two electrons ({\it b}) and single energy spectrum
of the electrons ({\it c}), after background
subtraction from $^{100}$Mo with of 7.369 kg y  
exposure \cite{ARN05}.
The solid line corresponds to the expected
spectrum from $2\nu\beta\beta$ simulations and the shaded
histogram is the subtracted background
computed by Monte-Carlo simulations.}
\end{center}
\end{figure}

\begin{table}
\caption{Investigated isotopes with NEMO-3 \cite{ARN05a}}
\vspace{0.5cm}
\begin{center}
\begin{tabular}{c|c|c|c|c|c|c|c}
\hline
Isotope & $^{100}$Mo & $^{82}$Se & $^{130}$Te & $^{116}$Cd &
$^{150}$Nd & $^{96}$Zr & $^{48}$Ca  \\
\hline
Enrichment, & 97 & 97 & 89 & 93 & 91 & 57 & 73 \\
\% & & & & & & & \\
Mass of & 6914 & 932 & 454 & 405 & 36.6 & 9.4 & 7.0 \\
isotope, g & & & & & & & \\
\hline
\end{tabular}
\end{center}
\end{table}

Fig. 4 and Fig. 5 display the
spectrum of $2\nu\beta\beta$ events
for $^{100}$Mo and $^{82}$Se that were collected over 389 days (Phase I)
\cite{ARN05}. For $^{100}$Mo the angular distribution
(Fig. 4{\it b}) and single electron spectrum (Fig. 4{\it c}) are also shown.
The total number of events exceeds 219,000 which is much greater
than the total statistics of all of the preceding
experiments with $^{100}$Mo (and even greater than the total statistics of all previous 
$2\nu\beta\beta$ decay experiments!). It should also be noted that the
background is as low as $2.5\%$ of the total number of
$2\nu\beta\beta$ events. Employing the calculated values of the detection
efficiencies for
$2\nu\beta\beta$ events, the following half-life values were
obtained for $^{100}$Mo and $^{82}$Se \cite{ARN05}:

\begin{equation}
T_{1/2}(^{100}{\rm Mo};2\nu) = [7.11 \pm 0.02({\rm stat.}) \pm 0.54({\rm syst.})    
]\times 10^{18}\; y,                     \
\end{equation}

\begin{equation}
T_{1/2}(^{82}{\rm Se};2\nu) = [9.6 \pm 0.3({\rm stat.}) \pm 1.0({\rm syst.}) ]\times
10^{19}\; y.\
\end{equation}

\begin{figure}
  \begin{center} 
    \includegraphics[height=10cm]{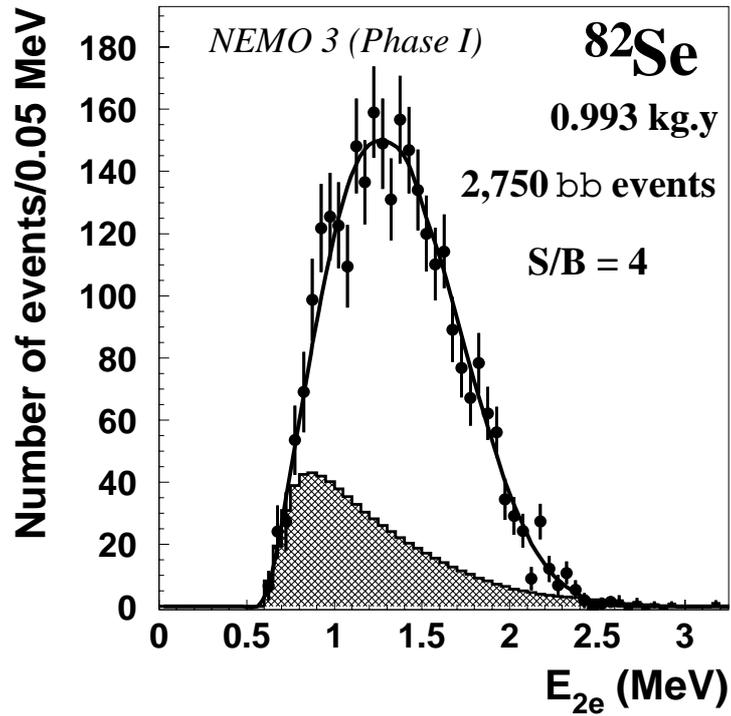} 
  \end{center}
  \caption{Energy sum spectrum of the two
electrons after background subtraction
from $^{82}$Se with 0.993 kg y exposure (same legend as
Fig. 4) \cite{ARN05}. The signal contains
 2,750 2$\beta$ events and the signal-to-background ratio is
4.} 
  \label{figure5}  
\end{figure} 

These results and results for $^{48}$Ca, $^{96}$Zr, $^{116}$Cd, $^{130}$Te and $^{150}$Nd
are presented in Table 8. Notice that the values for
$^{100}$Mo and $^{116}$Cd have been obtained on the assumption
that the single state dominance (SSD) mechanism is valid\footnote{Validity of SSD mechanism
in $^{100}$Mo was demonstrated using analysis of the single electron 
spectrum (see \cite{ARN04,SHI06}). In the case of $^{116}$Cd this is still 
a hypothesis.} \cite{SIM01a,DOM05}.
Systematic uncertainties can be decreased using
calibrations and can be improved by up to $\sim$ (3--5)$\%$.

\begin{table}
\caption{Two neutrino half-life values for different nuclei
obtained in
the NEMO-3 experiment (for $^{116}$Cd, $^{96}$Zr
$^{150}$Nd, $^{48}$Ca and $^{130}$Te the results are preliminary). 
First error is statistical and second is systematic; 
{\it S}/{\it B} is the signal-to-
background ratio.}
\vspace{0.5cm}
\begin{center}
\begin{tabular*}{\textwidth}{l|@{\extracolsep{\fill}}c|c|c|c}
\hline
Isotope & Measurement & Number of & {\it S}/{\it B} & $T_{1/2}(2\nu)$, y \\
& time, days & $2\nu$ events & &  \\
\hline
$^{100}$Mo & 389 & 219000 & 40 & $(7.11 \pm 0.02 \pm
0.54 )\times 10^{18}$ \cite{ARN05}  \\
$^{82}$Se & 389 & 2750 & 4 & $(9.6 \pm 0.3 \pm 1.0)
\times 10^{19}$ \cite{ARN05} \\
$^{116}$Cd & 1222 & 6949 & 10 & $(2.88 \pm 0.04 \pm
0.16)\times 10^{19}$  \\
$^{96}$Zr & 1221 & 428 & 1 & $(2.35 \pm 0.14 \pm 0.19)
\times 10^{19}$ \cite{ARN10} \\
$^{150}$Nd & 939 & 2018 & 2.8 & $(9.2^{+0.25}_{-0.22} \pm 0.62)
\times 10^{18}$ \cite{ARG09} \\
$^{48}$Ca & 943.16 & 116 & 6.8 & $(4.4^{+0.5}_{-0.4} \pm 0.4)
\times 10^{19}$  \\
$^{130}$Te & 1152 & 236 & 0.35 & $(6.9 \pm 0.9 \pm 1.0)
\times 10^{20}$  \\
\hline
\end{tabular*}
\end{center}
\end{table}

Figure 6 shows the tail of the two-electron energy sum spectrum in
the $0\nu\beta\beta$ energy window
for $^{100}$Mo and $^{82}$Se (Phase I+II; 3.75 yr of measurement).
One can see that the experimental spectrum is in good agreement with
the calculated spectrum, which was obtained taking
into account all sources of background. Using a maximum
likelihood method, the following
limits on neutrinoless double beta decay of $^{100}$Mo and
$^{82}$Se (mass mechanism; 90\% C.L.) have been obtained:

\begin{equation}
T_{1/2}(^{100}{\rm Mo};0\nu) > 1.1\times 10^{24}\; {\rm y},\
\end{equation}

\begin{equation}
T_{1/2}(^{82}{\rm Se};0\nu) > 3.6\times 10^{23}\; {\rm y}.\
\end{equation}

\begin{figure}
\begin{center}
\includegraphics[scale=0.5]{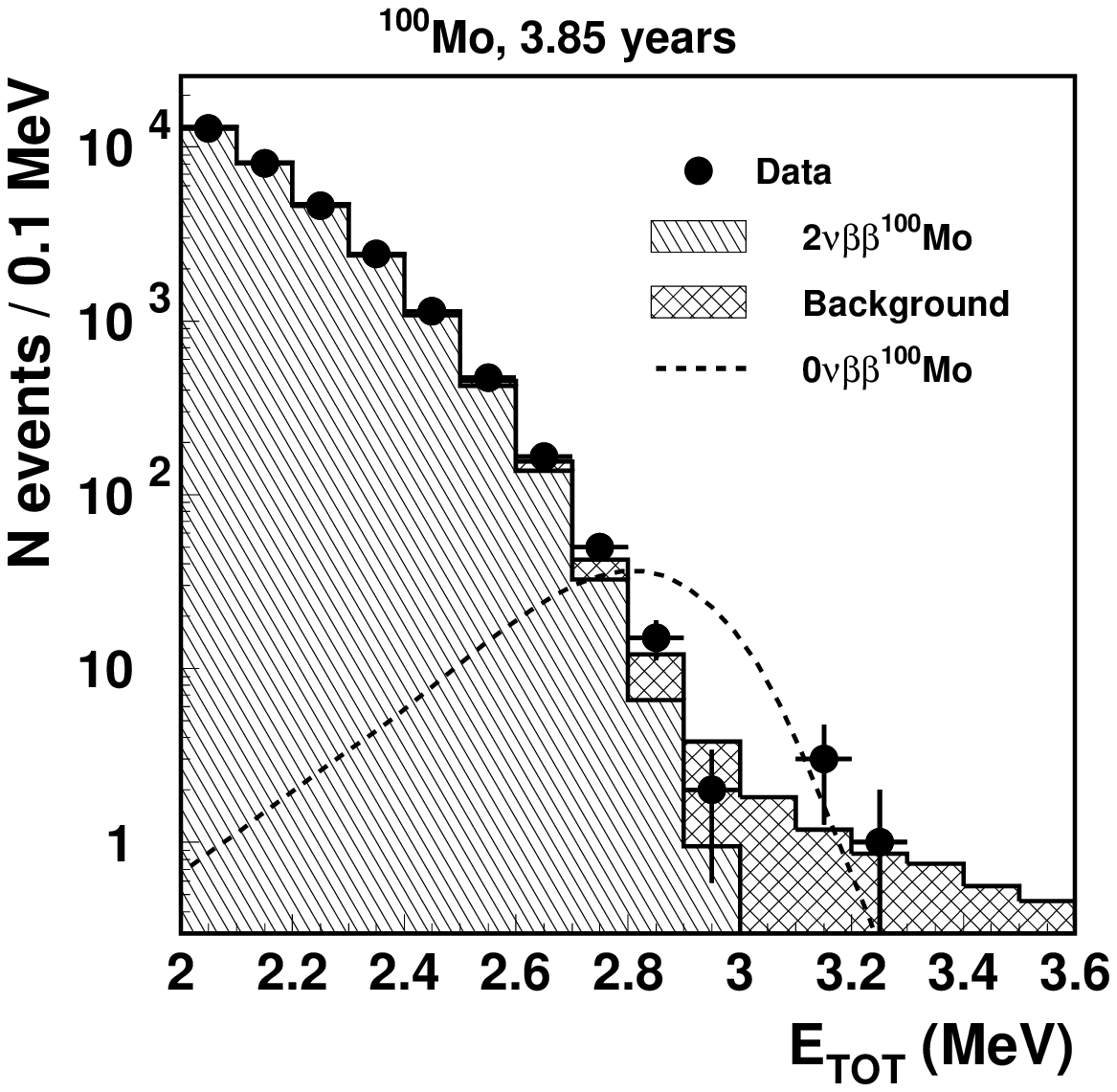}
\includegraphics[scale=0.5]{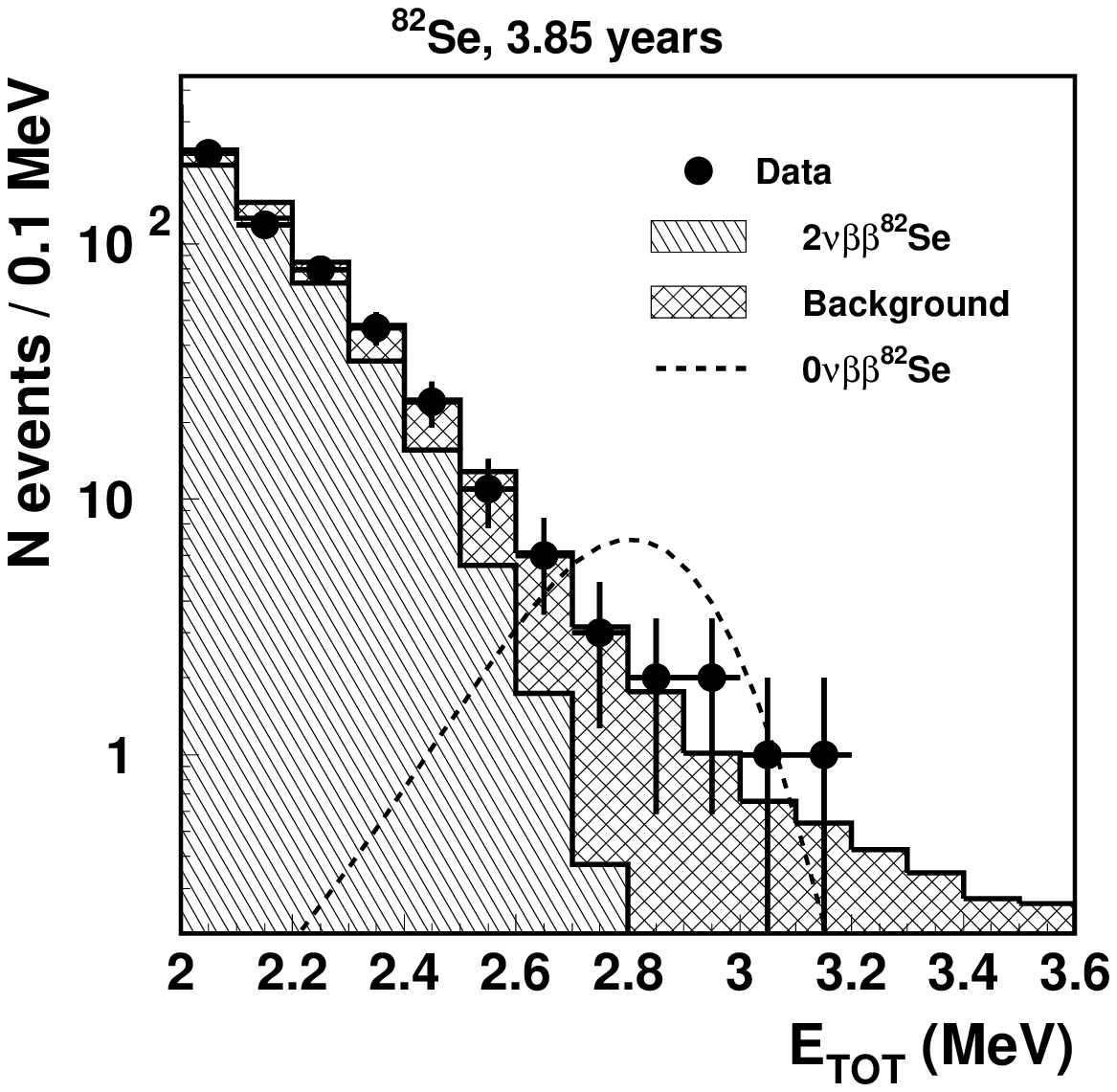}
\caption{\label{fig:figure6} Distribution of the energy sum of two electrons 
in the region around Q$_{\beta\beta}$ value 
for $^{100}$Mo (a) and  $^{82}$Se (b), 1409 d data \cite{BAR10a}.
High energy tail 
of the energy sum distribution for events in
molybdenum (left) and selenium (right) foils are shown with black
points. The background contributions are shown within the
histogram. The shape of a hypothetical $0\nu$ signal  is shown by the 
curve in arbitrary units.}
\end{center}
\end{figure}

Additionally, using NME values from \cite{ROD06,KOR07,KOR07a,SIM08} the bound on
$\langle m_{\nu} \rangle$ gives 0.45--0.93 eV for $^{100}$Mo
and 0.89--1.61 eV for $^{82}$Se.

In this experiment the best present limits on all possible modes
of double beta decay with Majoron emission
have been obtained too (see Tables 3 and 4). 

NEMO-3 experiment will be running up to end of 2010.

\section{Planned experiments}

Here seven of the most developed and 
promising experiments which can
be realized within the next few years are discussed 
(see Table 9). The
estimation of the sensitivity in the experiments is made using
NMEs
from \cite{ROD06,KOR07,KOR07a,SIM08,CAU08}.

\begin{table}
\caption{Seven most developed and promising projects. 
Sensitivity at 90\% C.L. for three (1-st step of GERDA and MAJORANA, SNO+, and KamLAND-Xe) 
five (EXO, SuperNEMO and CUORE) and ten (full-scale GERDA and MAJORANA) 
years of measurements is presented. M - mass of isotopes.}
\vspace{0.5cm}
\begin{center}
\begin{tabular}{c|c|c|c|c|c}
\hline
Experiment & Isotope & M, kg & Sensitivity & Sensitivity & Status \\
& &  & $T_{1/2}$, y & $\langle m_{\nu} \rangle$, meV &  \\
\hline
CUORE  & $^{130}$Te & 200 & $6.5\times10^{26}$$^{*)}$ & 20--50 & in progress \\ 
\cite{ARNA04,SIS10} & & & $2.1\times10^{26}$$^{**)}$ & 35--90 & \\
GERDA \cite{ABT04} & $^{76}$Ge & 40 & $2\times10^{26}$ & 70--300 & in progress \\
& & 1000 & $6\times10^{27}$ & 10--40 & R\&D\\ 
MAJORANA & $^{76}$Ge & 30--60 & (1--2)$\times10^{26}$ & 70--300 & R\ in progress \\
\cite{MAJ03, AVI10}& & 1000 & $6\times10^{27}$ & 10--40 & R\&D \\ 
EXO \cite{DAN00} & $^{136}$Xe & 200 & $6.4\times10^{25}$ & 95--220 & in progress \\
& & 1000 & $8\times10^{26}$ & 27--63 & R\&D \\ 
SuperNEMO & $^{82}$Se & 100--200 & (1--2)$\times10^{26}$ & 40--110 & in progress \\
\cite{BAR02,BAR04a,SAA09} & & & & &\\
KamLAND-Xe & $^{136}$Xe & 400 & 4.5$\times10^{26}$ & 40--80 & in progress \\
\cite{NAK10} & & & & &\\
SNO+ \cite{KRA10} & $^{150}$Nd & 56 & $\sim$ 4.5$\times10^{24}$ & 100--300 & in progress \\
 & & 500 & $\sim$ 3$\times10^{25}$ & 40-120 &  R\&D \\
\hline
\end{tabular}
\flushleft{$^{*)}$ For the background 
0.001 keV$^{-1}$ kg$^{-1}$ y$^{-1}$; $^{**)}$ for the background 
0.01 keV$^{-1}$ kg$^{-1}$ y$^{-1}$.}
\end{center}
\end{table}

\subsection{CUORE \cite{ARNA04,SIS10}}

This experiment will be
run at the Gran Sasso Underground Laboratory (Italy; 3500 m w.e.). 
The plan is to
investigate 760 kg of $^{\rm nat}$TeO$_{2}$ , with a total of
$\sim$ 200 kg of $^{130}$Te. One thousand low-temperature ($\sim$ 8 mK)
detectors, each having a weight of 750 g, will be
manufactured and arranged in 19 towers. One tower is approximately
equivalent to the CUORICINO detector \cite{AND11}. 
Planed energy resolution is 5 kev (FWHM).
 One of the problems here is to reduce the
background level by a factor of about 10
to 100 in relation to the background level achieved in the
detector CUORICINO. Upon reaching a
background level of 0.001 keV$^{-1}$ kg$^{-1}$ y$^{-1}$,
the sensitivity of the experiment to
the $0\nu$ decay of $^{130}$Te for 5 y of measurements and at
90\% C.L. will
become approximately
$6.5 \times 10^{26}$ y ($\langle m_{\nu} \rangle$ $\sim$ 0.02--0.05
eV). For more realistic level of background  
0.01 keV$^{-1}$ kg$^{-1}$ y$^{-1}$ 
sensitivity will be $\sim 2.1\times 10^{26}$ y for half-life and 
$\sim$ 0.04--0.09 eV for the effective Majorana neutrino mass. 
The experiment has been approved and funded. A general test of the CUORE detector, 
comprising a single tower and
named CUORE-0, will take data in 2011.

\subsection{GERDA \cite{ABT04}}

This is one of two  planned experiments
with $^{76}$Ge (along with the MAJORANA experiment). The experiment is to be
located in the Gran Sasso Underground Laboratory (Italy, 3500 m w.e.). The
proposal is based on ideas and approaches which
were proposed for GENIUS \cite{KLA98} and the GEM \cite{ZDE01}
experiments.
The plan is to place ``naked'' HPGe detectors in highly
purified liquid argon (as passive and active shield). It minimizes the weight of construction
material near the detectors and
decreases the level of background. The liquid argon dewar is
placed into a vessel of very pure water.
The water plays a role of passive and active (Cherenkov radiation)
shield.

The proposal involves three phases. In the first phase, the
existing HPGe detectors ($\sim$ 18 kg), which
previously were used in the Heidelberg-Moscow \cite{KLA01} and IGEX
\cite{AAL02a} experiments, will be utilized. In the second phase
$\sim$ 40 kg of enriched Ge will be investigated. In the third
phase the plan
 is to use $\sim$ 500--1000 kg of $^{76}$Ge.

The first phase, lasting one year, is to measure with a sensitivity of
$3 \times 10^{25}$ y, that gives a possibility
of checking the ``positive'' result of \cite{KLA04,KLA06,KLA01a}. The
sensitivity of the second phase (for three years of measurement)
will be $\sim$ $2 \times 10^{26}$ y. This corresponds to a
sensitivity for $\langle m_{\nu} \rangle$ at the level of
$\sim$ 0.07--0.3 eV.

The first two phases have been approved and funded. The first phase 
set-up is in an advanced construction stage and data
taking is foreseen for 2011.
The results of this
first step will play an important role in the decision to support
the full scale experiment.

The project is very promising although it will be difficult to
reach the desired level of background.
One of the significant problems is $^{222}$Rn in the liquid
argon (see, for example, results of \cite{KLAP04}) and, in addition, 
background from $^{42}$Ar can be a problem too.

\subsection{MAJORANA \cite{MAJ03, AVI10}}

The MAJORANA facility will consist of $\sim$ 1000 HPGe detectors
manufactured from enriched germanium (the degree
of enrichment is $>$ 86\%). The total mass of enriched germanium
will be 1000 kg. The facility is designed in such a
way that it will consist of many individual supercryostats
manufactured from low radioactive copper, each containing
HPGe detectors. The entire facility will be surrounded by a
passive shield and will be located at an underground
laboratory in the United States.
Only the total energy deposition will be utilized in measuring the
$0\nu\beta\beta$ decay of $^{76}$Ge to the ground
state 
of the daughter nucleus. The use of HPGe detectors,
pulse shape analysis, anticoincidence,
and low radioactivity structural materials will make it possible
to reduce
the background to a value below 
$3 \times 10^{-4}$ keV$^{-1}$ kg$^{-1}$ y$^{-1}$  and to reach
a sensitivity of about $6 \times 10^{27}$ y within
ten years of measurements. The corresponding sensitivity to the
effective mass of the Majorana neutrino is about
0.01 to 0.04 eV.
The measurement of the $0\nu\beta\beta$ decay of $^{76}$Ge to the
0$^{+}$ excited state of the daughter nucleus
will be performed
by recording two cascade photons and two beta electrons. The
planned sensitivity for this process is about 10$^{27}$ y.

In the first step $\sim$ 30--60 kg of $^{76}$Ge will be
investigated. It is anticipated that the sensitivity to
$0\nu\beta\beta$ decay to the ground state of the daughter nuclei
for 3 years of measurement will be (1--2)$\times 10^{26}$ y.
It will reject or to confirm the ``positive'' result from
\cite{KLA04,KLA06,KLA01a}. Sensitivity to
$\langle m_{\nu} \rangle$ will be $\sim$ 0.07--0.3 eV. During this
time different methods and technical
questions will be checked and possible background problems will be
investigated. The first module of MAJORANA (DEMONSTRATOR) is under constriction now
and measurements is planned to begin in 2013.

\subsection{EXO \cite{DAN00}}

In this experiment the plan is to implement Moe's proposal of
1991 \cite{MOE91}. Specifically it is to record both ionization
electrons
and the Ba$^{+}$ ion originating from the double-beta-decay
process $^{136}$Xe$\to$$^{136}Ba^{++}$ + 2e$^{-}$. In \cite{DAN00}, it is
proposed to
operate with 1t of $^{136}$Xe. The actual technical implementation
of
the experiment has not yet been
developed. One of the possible schemes is to fill a TPC
with liquid enriched xenon. To avoid the background from the
2$\nu$ decay of $^{136}$Xe, the
energy resolution
of the detector must not be poorer than 3.8\% (FWHM) at an energy
of 2.5 MeV (ionization and scintillation signals
will be detected).

In the 0$\nu$ decay of $^{136}$Xe, the TPC will measure the energy
of two electrons and the coordinates of the event
to within a few millimeters. After that, 
using a special stick Ba ions will be removed from the liquid and then 
will be registered in a special cell 
by resonance excitation. For Ba$^{++}$ to undergo
a transition to a state of Ba$^{+}$, a special gas is added to
xenon. The authors of the project assume that the
background will be reduced to one event within five years of
measurements. Given a 70\% detection efficiency
it will be possible to reach a sensitivity of about $8 \times
10^{26}$ y for the $^{136}$Xe half-life and a
sensitivity of about 0.03 to 0.06 eV for the neutrino mass.

One should note that the principle difficulty in this experiment
is associated with detecting the
 Ba$^{+}$ ion with a reasonably high efficiency.
This issue calls for thorough experimental tests, and positive
results have yet to be obtained.

As the first stage of the experiment EXO-200
will use 200 kg of $^{136}$Xe without Ba ion identification.
This experiment is currently
under preparation and measurements will start probably in
2011.
The 200 kg of enriched Xe is a product of
Russia with an enrichment of $\sim 80\%$. If the background is 
40 events in 5 y of measurements,
 as estimated by the
authors, then the sensitivity of the experiment will be $\sim$
$6 \times 10^{25}$ y. This corresponds to sensitivity for $\langle
m_{\nu} \rangle$ at the level $\sim$ 0.1--0.2 eV. This initial prototype 
will operate at the Waste Isolation Pilot Plant (WIPP) in Southern 
New Mexico (USA).

\subsection{SuperNEMO \cite{BAR02,BAR04a,SAA09}}

The NEMO Collaboration has studied and is pursing an experiment
that will observe 100--200 kg of $^{82}$Se
with the aim of reaching a sensitivity for the $0\nu$ decay mode at
the level of
$T_{1/2} \sim$ (1--2) $\times 10^{26}$ y.
The corresponding sensitivity to the neutrino mass is 0.04
to 0.11 eV. In order to accomplish this
goal, it is proposed to use the experimental procedures nearly
identical to that in the NEMO-3 experiment
(see Subsection 3.1). The new detector will have planar geometry
and will consist of 20
identical modules (5-7 kg of $^{82}$Se in each sector). A $^{82}$Se
source having a thickness of
about 40 mg/cm$^{2}$ and a very low content of radioactive
admixtures is placed at the center of
the modules. The detector will again record all features of double
beta
decay: the electron energy will be recorded by counters based on
plastic scintillators ($\Delta E/E \sim$ 8--10$\% (FWHM) $
at {\it E} = 1 MeV),
while tracks will be reconstructed with the aid of Geiger
counters.
The same device can be used to investigate $^{150}$Nd, $^{100}$Mo, $^{116}$Cd,
and $^{130}$Te with a sensitivity to
$0\nu\beta\beta$ decay at a
level of about (0.5--1) $\times 10^{26}$ y.

The use of an already tested experimental technique is an
appealing feature of this experiment. The plan is
to arrange the equipment at the new Frejus Underground Laboratory
(France; the respective depth being 4800 m w.e.).
The experiment is currently in its  R\&D stage. The construction and commissioning
of the demonstrator (first module) will be completed in 2013.

\subsection{SNO+}

SNO+ is an upgrade of the solar neutrino experiment SNO (Canada),
aiming at filling the SNO detector with Nd-loaded liquid scintillator to investigate
the isotope $^{150}$Nd \cite{KRA10}. The present plan is to use 0.1\% natural Nd-loaded liquid
scintillator in 1000 tones, providing a source of 56 kg $^{150}$Nd. SNO+ is in construction 
phase with natural neodymium. Data taking is foreseen in 2012. After 3 yr of data tacking 
sensitivity will be $\sim 4.5\times10^{24}$ yr (or ~ 0.1-0.3 eV for $\langle m_{\nu} \rangle$). 
Finally 500 kg of enriched 
$^{150}$Nd will be used (if enrichment of such quantity of Nd will be possible). Planned sensitivity 
is $\sim 3\times10^{25}$ yr.

\subsection{KamLAND-Xe}

KamLAND-Xe is an upgrade of the KamLAND set-up \cite{NAK10}. The idea is to convert it to neutrinoless 
Double Beta Decay search by dissolving Xe gas in the liquid scintillator. This approach 
was proposed by R. Raghavan in 1994 \cite{RAG94}. This mixture (400 kg of Xe in 16 tons of liquid scintillator) 
will be contained in a small balloon suspended in the centre of the KamLAND sphere. 
It will guarantee low background level and high sensitivity of this experiment (see Table 9). 
The program should
start in 2011 (first phase) with 400 kg of isotope and continue in 2013 (2015) with 1 ton of
Xenon enriched to 90\% in $^{136}$Xe.

\section{Conclusions}

In conclusion, two-neutrino double-beta decay has so far been
recorded for ten nuclei ($^{48}$Ca, $^{76}$Ge,
$^{82}$Se, $^{96}$Zr, $^{100}$Mo,
$^{116}$Cd, $^{128}$Te, $^{130}$Te, $^{150}$Nd, $^{238}$U). In
addition, the $2\beta(2\nu)$ decay of $^{100}$Mo
 and $^{150}$Nd to the 0$^{+}$ excited state of the daughter
nucleus has been observed and the ECEC(2$\nu$) process in
$^{130}$Ba was observed. Experiments studying two-neutrino double
beta decay are presently approaching a qualitatively new level,
where high-precision measurements are performed
not only for half-lives but also for
all other parameters of the process.
As a result, a trend is emerging toward thoroughly investigating
all aspects of two-neutrino double-beta decay,
and this will furnish very important information about the values
of NME, the parameters
of various theoretical models, and so on. In this connection, one
may expect advances in the calculation of
NME and in the understanding of the nuclear
physics aspects of double beta decay.

Neutrinoless double beta decay has not yet been confirmed. There
is
a conservative limit on the effective value
of the Majorana neutrino mass at the level of 0.75 eV. 

The next-generation experiments, where the mass of the isotopes
being studied
will be as grand as 100 to 1000 kg, will have started within a few years. In all probability,
they will make it possible to reach the sensitivity for the
neutrino mass at a level of 0.01 to 0.1 eV. First step of GERDA (18 kg of $^{76}$Ge), 
EXO-200 (200 kg of $^{136}$Xe, CUORE-0 ($\sim$ 40 kg of natural Te) 
and KamLAND-Xe (400 kg of $^{136}$Xe) plan to start data-tacking in 2011.

\section{Acknowledgements}

This work was supported by the State Atomic Energy Corporation ROSATOM.

\end{document}